\documentclass[%
preprint,
superscriptaddress,
amsmath,amssymb,
prl,
prstper,
floatfix,
]{revtex4-1}

\usepackage{graphicx}% Include figure files
\usepackage{dcolumn}% Align table columns on decimal point
\usepackage{bm}% bold math
\usepackage[dvipsnames]{xcolor}

\begin{document}

\title{Single and multi-mode directional lasing from arrays of dielectric nanoresonators}

\author{Shaimaa I. Azzam}
% \email{sazzam@purdue.edu}
\affiliation{School of Electrical \& Computer Engineering and Birck Nanotechnology Center, Purdue University, West Lafayette, Indiana, 47907, USA}
\affiliation{Purdue Quantum Science and Engineering Institute, Purdue University, West Lafayette, Indiana, 47907, USA}

\author{Krishnakali  Chaudhuri}
\affiliation{School of Electrical \& Computer Engineering and Birck Nanotechnology Center, Purdue University, West Lafayette, Indiana, 47907, USA}

\author{Alexei Lagutchev}
\affiliation{School of Electrical \& Computer Engineering and Birck Nanotechnology Center, Purdue University, West Lafayette, Indiana, 47907, USA}

\author{Zubin Jacob}
\affiliation{School of Electrical \& Computer Engineering and Birck Nanotechnology Center, Purdue University, West Lafayette, Indiana, 47907, USA}

\author{Young L. Kim}
\affiliation{Weldon School of Biomedical Engineering, Purdue University, West Lafayette, IN 47907, USA }

\author{Vladimir M. Shalaev}
\email{shalaev@purdue.edu}
\affiliation{School of Electrical \& Computer Engineering and Birck Nanotechnology Center, Purdue University, West Lafayette, Indiana, 47907, USA}
\affiliation{Purdue Quantum Science and Engineering Institute, Purdue University, West Lafayette, Indiana, 47907, USA}

\author{Alexandra Boltasseva} 
\email{aeb@purdue.edu}
\affiliation{School of Electrical \& Computer Engineering and Birck Nanotechnology Center, Purdue University, West Lafayette, Indiana, 47907, USA}
\affiliation{Purdue Quantum Science and Engineering Institute, Purdue University, West Lafayette, Indiana, 47907, USA}

\author{Alexander V. Kildishev}
\email{kildishev@purdue.edu}
\affiliation{School of Electrical \& Computer Engineering and Birck Nanotechnology Center, Purdue University, West Lafayette, Indiana, 47907, USA}
\affiliation{Purdue Quantum Science and Engineering Institute, Purdue University, West Lafayette, Indiana, 47907, USA}

\date{\today}
 
\begin{abstract}
The strong electric and magnetic resonances in dielectric subwavelength structures have enabled unique opportunities for efficient manipulation of light-matter interactions. 
Besides, the dramatic enhancement of nonlinear light-matter interactions near so-called bound states in the continuum (BICs) has recently attracted enormous attention due to potential advancements in all-optical and quantum computing. 
However, the experimental realizations and the applications of high-Q factor resonances in dielectric resonances in the visible regime have thus far been considerably limited. 
In this work, we explore the interplay of electric and magnetic dipoles in arrays of dielectric nanoresonators to enhance light-matter interaction. We report on the experimental realization of high-Q factor resonances in the visible through the collective diffractive coupling of electric and magnetic dipoles.  Providing direct physical insights, we also show that coupling the Rayleigh anomaly of the array with the electric and magnetic dipoles of the individual nanoresonators can result in the formation of different types of BICs. 
We utilize the resonances in the visible regime to achieve lasing action at room temperature with high spatial directionality and low threshold.  Finally, we experimentally demonstrate multi-mode, directional lasing and study the BIC-assisted lasing mode engineering in arrays of dielectric nanoresonators. We believe that our results enable a new range of applications in flat photonics through realizing on-chip controllable single and multi-wavelength micro-lasers.
\end{abstract}

\pacs{Valid PACS appear here}
\maketitle

%\tableofcontents

% \section{Introduction}
Bound states in the continuum (BICs) are dark states that stay localized even though they coexist with the structure's radiation continuum \cite{von1929some, lee2012observation, plotnik2011experimental, moiseyev2009suppression, friedrich1985interfering, hsu2016bound, alu2018experimental, hsu2013observation}. Ideal BICs have infinite lifetime and require at least one dimension of the structure to extend to infinity \cite{hsu2016bound}. In practical realizations, due to perturbations and finite extent, the BIC collapses to a Fano resonance with a limited (yet still very long) lifetime -- a regime known as quasi-BIC \cite{hsu2016bound, sadrieva2017transition}. 
Besides, in the vicinity of the BIC, near-BIC high-Q resonances can be attained by slightly detuning the system from the BIC point \cite{azzam2018formation, fonda1961resonance, fonda1963bound}. Near-BIC resonances are generally stable to fabrication imperfections as they can be obtained at a relatively broad interval of a system parameter values compared to BICs, which are typically achieved at a single point \cite{ fonda1961resonance, fonda1963bound}. 
Photonic structures supporting BICs have been intensively studied over the last decade as they offer unique ways for engineering and enhancing the light-matter interactions \cite{marinica2008bound, hsu2016bound, alu2018experimental, hsu2013observation, azzam2018formation, kivshar2017high, alu2018embedded, Sadreev2008bound}. Photonic BICs have to date enabled a plethora of applications, including light routing \cite{yu2019photonic}, nonlinear enhancement \cite{koshelev2020subwavelength}, sensing \cite{leitis2019angle}, and lasing \cite{kante2017lasing, Arseniy2018, huang2020ultrafast}. 
However, the experimental realizations of BICs and their applications in the visible wavelength range have been considerably limited and started to gain interest only recently \cite{huang2020ultrafast}.

The emergence of dielectric nanoresonators with strong electric and magnetic responses has enabled fundamentally new ways of engineering light-matter interactions \cite{kuznetsov2016optically, kivshar2017nature_review}. The interplay between electric and magnetic resonances in a dielectric nanoresonator enables control over the direction of scattering, along with the spectral position, strength, and quality factor of the resonance \cite{van2013designing}. This control can be realized in a single scatterer by tuning its shape to support a desired electromagnetic response \cite{evlyukhin2011multipole}.
Additionally, arranging the nanoresonators in a lattice can influence their spectral response providing more degrees of control over %for the manipulation of% 
light-matter interaction \cite{babicheva2018lattice, li2018engineering}. 
 Coupling between the electric dipole (ED) and magnetic dipole (MD) in individual dielectric resonators can be controlled by placing the resonators in close proximity to engage their near field interaction. However, due to the high refractive index of dielectric nanoresonators' materials, the electromagnetic field is generally tightly confined inside the resonator and near field coupling is relatively low. 
 Collective diffractive coupling of EDs and MDs has been proposed as a more effective way of controlling the response of dielectric nanoresonators arrays \cite{li2018engineering}. This can be achieved through the coupling of ED and MD resonances in an array of dielectric scatterers to the Rayleigh anomaly of the array. 

In this work, we investigate the experimental realization of high-Q factor coupled ED and MD resonances supported in the visible range by the array of dielectric nanoresonators due to their collective interaction.
Furthermore, we demonstrate lasing action within the visible spectrum, relying on the high-Q resonances supported by arrays of nanoresonators. The laser operates at room temperature with high beam directionality and low-threshold. 
In addition, we experimentally investigate the BIC-assisted mode engineering in arrays of dielectric nanoresonators, and we show multi-wavelength directional lasing. Two- and three-mode lasers have been demonstrated experimentally, showing the potential for this type of structure to support controllable multi-wavelength on-chip microlasers.

\section{Results}
\begin{figure*}[t] \centering
     \includegraphics[scale= 1]{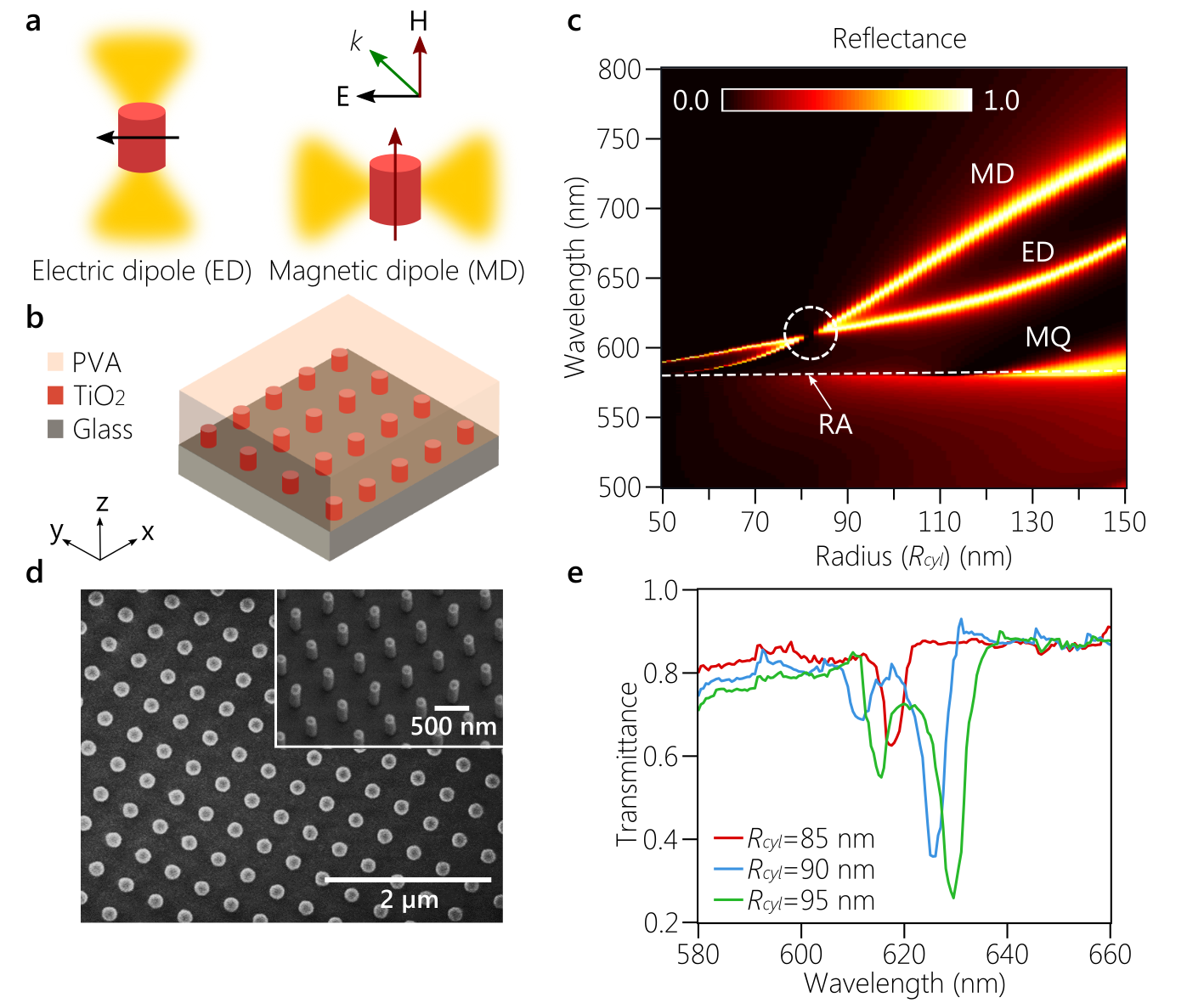}
     \caption{High-Q resonances in arrays of dielectric nanoresonators with electric and magnetic dipoles. a) Electric and magnetic dipole resonances supported by a dielectric nanoresonator. b) A schematic of the array structure with a square lattice of 400 nm period made of titanium oxide ($\text{TiO}_2$) cylindrical nanoresonators with a height of 300 nm on a glass substrate and covered with polyvinyl alcohol (PVA). c) Simulated reflectance as a function of the nanoresonator radius ($R_{cyl}$). The dashed white circle shows the disappearance of the resonance, indicating the realization of the Kerker condition at $R_{cyl}$ = 82 nm and $\lambda$ = 610 nm. The three resonances correspond to the magnetic dipole (MD), electric dipole (ED) and the magnetic quadrupole (MD), respectively and the dashed white line shows the Rayleigh anomaly (RA) of the array. d) An SEM image of the fabricated structure. e) The measured transmittance of the array  with radii of 85, 90, and 95 nm.}
\label{lasing_fig1} 
\end{figure*}

Two orthogonal orientations of dipole modes are generally supported in a dielectric resonator; a horizontal (electric dipole, ED) and vertical  (magnetic dipole, MD) (Fig. \ref{lasing_fig1}a). Here, we study an array comprising $\text{TiO}_2$ cylindrical nanoresonators with a height of 300 nm arranged in a square lattice with a period of 400 nm on a silica glass substrate, Fig. \ref{lasing_fig1}b. The structure is covered with a polymer, index-matched to the substrate. Our choice of materials and dimensions are to achieve resonances in the visible wavelength range.

Figure \ref{lasing_fig1}c shows the reflectance spectra of the array as a function of the radius of the nanoresonators. Three resonances corresponding to the ED, MD as well as a magnetic quadrupole (MQ) can be observed in the reflectance spectra, as indicated in Fig. \ref{lasing_fig1}c. 
The resonance classification is accomplished based on the displacement current loops and E-field profiles; more information can be found in Section 1 of Supplementary Materials (SM) \cite{SM}. 
The Rayleigh anomaly (RA) of the array is depicted as a dashed white line in Fig.\ref{lasing_fig1}c.
As the radius of the nanoresonators decreases, the linewidth of the MD resonance gets narrower and the position of the MD approaches that of the Rayleigh anomaly. Both observations are distinctive features of the diffractive coupling \cite{li2018engineering}. We also notice that the spectral position of the MQ follows the Rayleigh anomaly closely. 
The ED and MD modes spectrally overlap a radius of 82 nm, and a wavelength of 610 nm, highlighted by the dashed circle in Fig. \ref{lasing_fig1}c confirming the realization of Kerker's condition. Almost zero light reflection is realized at the Kerker point with a near-unity unidirectional transmission. 
Another feature that results from the diffractive coupling of the dipoles to the Rayleigh anomaly is the enhancement of electromagnetic fields (Fig. S2) which is very advantageous in boosting the nonlinear phenomenon.
More insight can also be gained from studying the effect of the nanoresonator height on the spectral positions and widths of the ED, MD, and MQ in Section 2 of the SM \cite{SM}. 

A scanning electron microscope (SEM) image of the fabricated structure is given in Fig. \ref{lasing_fig1}d. For details of the fabrication, see Methods and Fig. S5. Transmittance at normal incidence from a few samples with different $R_{cyl}$ was characterized using spectroscopic ellipsometry, Fig. \ref{lasing_fig1}e. A single resonance is observed at a radius value of 85 nm where the ED and MD start to overlap, as can be noticed from  Fig. \ref{lasing_fig1}e. 
At larger radii values, 90 nm and 95 nm, two dips corresponding to the ED and MD are observed. 
The measured transmittance is in a very good agreement with the simulated spectra, Fig. S4. However, due to the non-uniformity in the radii of the fabricated sample, the measured resonances are broadened, and their dip positions are slightly shifted compared to the simulated ones with around 5 nm.

\subsection{Lasing in arrays of dielectric nanoresonators}
After the successful demonstration of the passive nanoresonator arrays with high-Q factor resonances, we employ them to provide feedback for lasing action in the visible spectrum. First, the index matching coating is washed off. Then, the array is spin-coated with polyvinyl alcohol (PVA) solution of an organic gain medium (Rhodamine 101) with a concentration of 10 mM, which acts as an active material. The spin coating yields an active layer with a thickness of $\sim$ 2 $\mu \text{m}$. To characterize the lasing action, the sample is pumped with the second-harmonic output from an Nd: YAG laser (532 nm, 1-Hz repetition rate, and 400-ps pulse duration). The emitted light is collected through fiber and directed to a spectrometer to get the emission spectrum to demonstrate the two critical effects of lasing; threshold and spectral narrowing. The setup schematic is given in Fig. S7 and details of the experiment can be found in Methods.

\begin{figure*}[t] \centering
     \includegraphics[scale= 1.0]{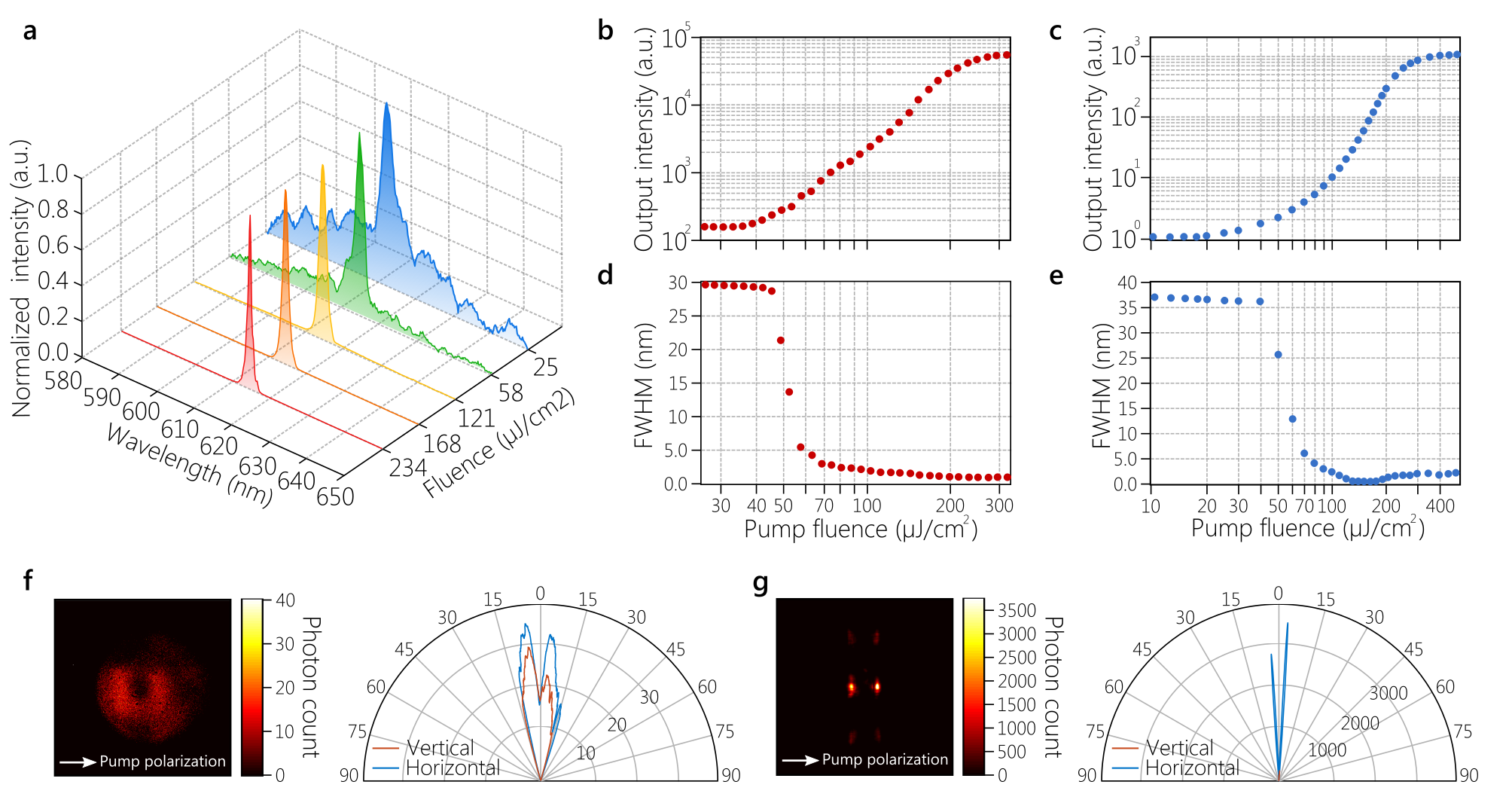}
          \caption{Lasing action with near-BIC feedback. a) The evolution of the emission spectra collected from the gain-coated structure with the increased pump fluence, showing a transition from amplified spontaneous emission to lasing. b, c) The experimental (b) and simulated (c) output intensities vs. the pump fluence. d, e) The experimental (d) full width at half-maximum (FWHM) and simulated (e) FWHM as a function of pump fluence. f, g) Back focal plane image and the radiation patterns, which are vertical and horizontal cross-sections of the focal plane image, of the laser emission below (f) and above (g) lasing threshold. White arrows in f, g show the polarization of the pump laser. }
\label{lasing_fig2} 
\end{figure*}

Figure \ref{lasing_fig2}a depicts the evolution of the collected emission spectra as the pump fluence increases showing a narrow lasing peak emerging around 614 nm. An apparent threshold behavior is observed where the emission transitions from enhanced spontaneous to stimulated emission at around 40 $\mu \text{J/cm}^2$, as evident from the experimental (Fig. \ref{lasing_fig2}b) and simulated (Fig. \ref{lasing_fig2}c) data. This process is accompanied by a significant narrowing of the emission spectral width from $>$ 30 nm to $<$ 1 nm, as observed in the experiment (Fig. \ref{lasing_fig2}d), and numerical simulations (Fig. \ref{lasing_fig2}e). The emission directionality is characterized using back focal plane measurements, see Methods. Below the lasing threshold, the enhanced spontaneous emission is illustrated with a back focal plane cross-section and the emission pattern of Fig. \ref{lasing_fig2}f with the cross-section showing a realtively wide spatial spread. However, above the threshold, the same measurements (see, Fig. \ref{lasing_fig2}g) indicate tightly focused spots and narrow beams at 3.5$^o$ with a less than 0.8$^o$ in beamwidth. It is worth mentioning that due to our structure's square lattice, the back focal plane image should consist of two points along the x-axis and two along the y-axis. However, Fig. \ref{lasing_fig2}f shows only a pair of bright spots along the x-axis, which is a result of the polarization of the pump. When an analyzer with orthogonal orientation (along the y-axis) is introduced in the lasing emission path before collection, we observe four bright spots in the back focal plane emission as expected, Fig. S8.
All our experimental studies are guided and supported by the advanced multiphysics numerical framework built on coupling carrier kinetics with a full-wave time-domain Maxwell equation solver \cite{azzam2019exploring}. Details of the numerical simulations can be found in Methods and Section 2 of SM \cite{SM}.

To the best of our knowledge, this is the first report of lasing action in the visible spectrum with feedback from an all-dielectric high-Q factor nanoresonator array. Unique features of the proposed system include a low threshold, tunable-by-design high-Q factor resonances, and room temperature operation with no need for complex cryogenic infrastructure. All these advantages suggest the suitability of the proposed laser for integration into a multitude of applications such as sensors, biological probes, and on-chip light sources.

\subsection{ Mode engineering near BIC states}
 
By engineering the geometrical and material parameters of a unit-cell dielectric resonator, we can selectively excite distinct resonant modes and also tune their spectral properties \cite{yang2014eit, campione2016broken}. Therefore, an  optimal unit-cell design can enable an excellent control over the band-diagram of the entire structure array. In general, band-diagram engineering can be accomplished through geometrical optimization which when applied to metasurfaces, results in complex shapes with astringent unit cell geometries and dimensions leading to further fabrication constraints \cite{wang2017band, yang2014eit, campione2016broken}. To perform mode engineering in our study, we intentionally limit the structure design to a simple, easy-to-fabricate square array of $\text{TiO}_2$ cylinders. A unit-cell of this type appears to be an ideal prototype for analyzing the interplay between the electric and magnetic responses in a dielectric nanoresonator, studying the array effect, as well as exploiting the intriguing physics of BICs. 

\begin{figure*}[t] \centering
     \includegraphics[scale= 1.2]{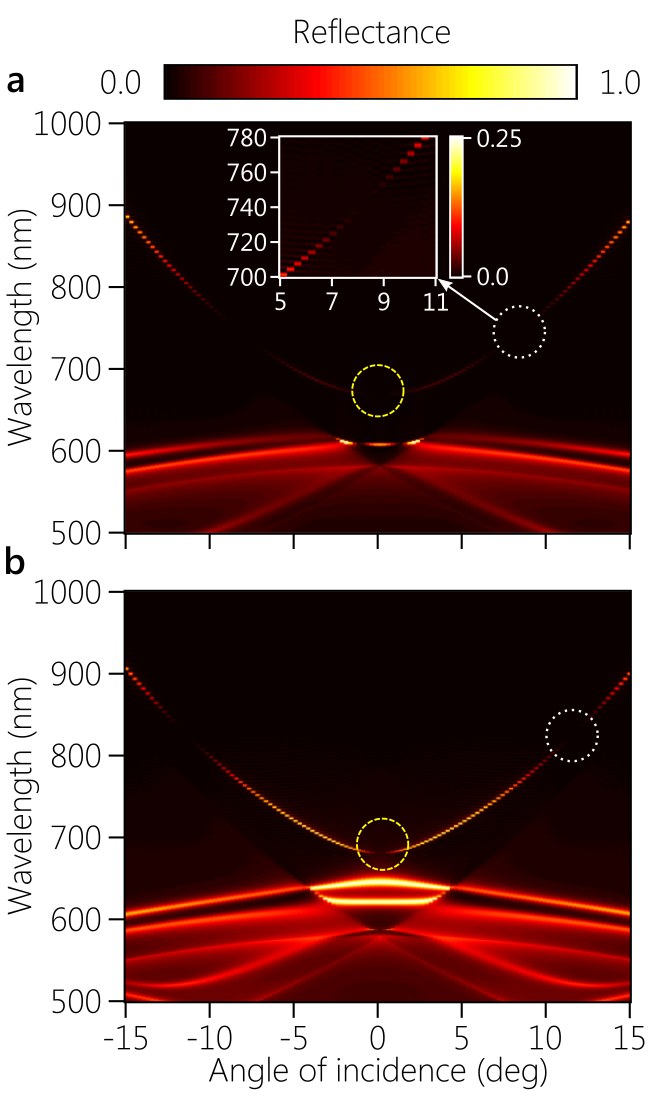}
     \caption{The simulated reflectance vs. the angle of incidence and wavelength. a) The reflectance of an array with $\text{TiO}_2$ cylinder nanoresonators ($R_{cyl}$ = 80 nm). The dashed yellow circle at the $\Gamma$ point highlights the symmetry-protected BIC at 669 nm. The dotted white circle highlights the off-$\Gamma$ (Friedrich-Wintgen, FW) BIC, which is zoomed in in the inset at 741 nm. b) An array with $R_{cyl}$ = 100 nm exhibiting the symmetry-protected BIC at the $\Gamma$ point and 683 nm (dashed yellow circle) and an FW BIC at 824 nm and 11.5$^o$ (dotted white circle).}
\label{lasing_fig3} 
\end{figure*}

Figure \ref{lasing_fig3}a shows the simulated reflectance of an array comprising nanoresonators with $R_{cyl}$ = 80 nm as a function of the angle of incidence and wavelength. The resonance at the $\Gamma$ point around 610 nm corresponds to the point where the ED and the MD start to overlap, which we used as a feedback for the single-mode lasing demonstrated in Fig. \ref{lasing_fig2}. 
It is worth noting that the array supports two types of BIC states, one at the $\Gamma$ point with wavelength 669.2 nm, which is a symmetry-protected state. Additionally, an off-$\Gamma$ Friedrich-Wintgen (FW) BIC is also supported at $8^o$  at a wavelength of 741 nm \cite{friedrich1985interfering, hsu2016bound}. The FW BIC is highlighted with a dotted white circle in Fig. \ref{lasing_fig3}a and magnified in the inset. This emphasizes the ability of an array with such a simple unit cell structure to support multiple types of resonances with different underlying physics. Not only can this structure support simple "band-edge" mode at the $\Gamma$ point, but it also provides access to the non-trivial FW BIC states that are not symmetry-protected and hence are more resilient to structural imperfections. 

The structure with $R_{cyl}$ = 100 nm shows a different reflectance response with two resonances observed at the $\Gamma$ point corresponding to the electric dipole at and the magnetic dipole resonances at 622 nm and 646.5 nm, respectively. As can be seen in Fig. \ref{lasing_fig3}b the array with $R_{cyl}$ = 100 nm supports a symmetry-protected BIC at 683 nm, inside the dashed yellow circle as well as an FW BIC at 824 nm and $11.5^o$ inside the dotted white circle. The stark difference between the optical response of the two designs confirms an excellent opportunity to engineer the band-edge and bandgap of metasurfaces by tuning their geometrical parameters. 

We use this concept to achieve two- and three-mode lasing from arrays of dielectric nanoresonators. On-chip multi-mode lasers are invaluable tools for many applications such as optical signal processing, quantum computing, wavelength multiplexing for telecommunication, and others. In this section, we explore the engineering of the various resonance in play in our structure to realize multi-mode lasing.
We design, fabricate and experimentally demonstrate dual- and triple-mode laser solely based on the optimal choice of the radius of the dielectric resonators in the BIC-type metasurface array.

\begin{figure*}[t] \centering
     \includegraphics[scale= 1.00]{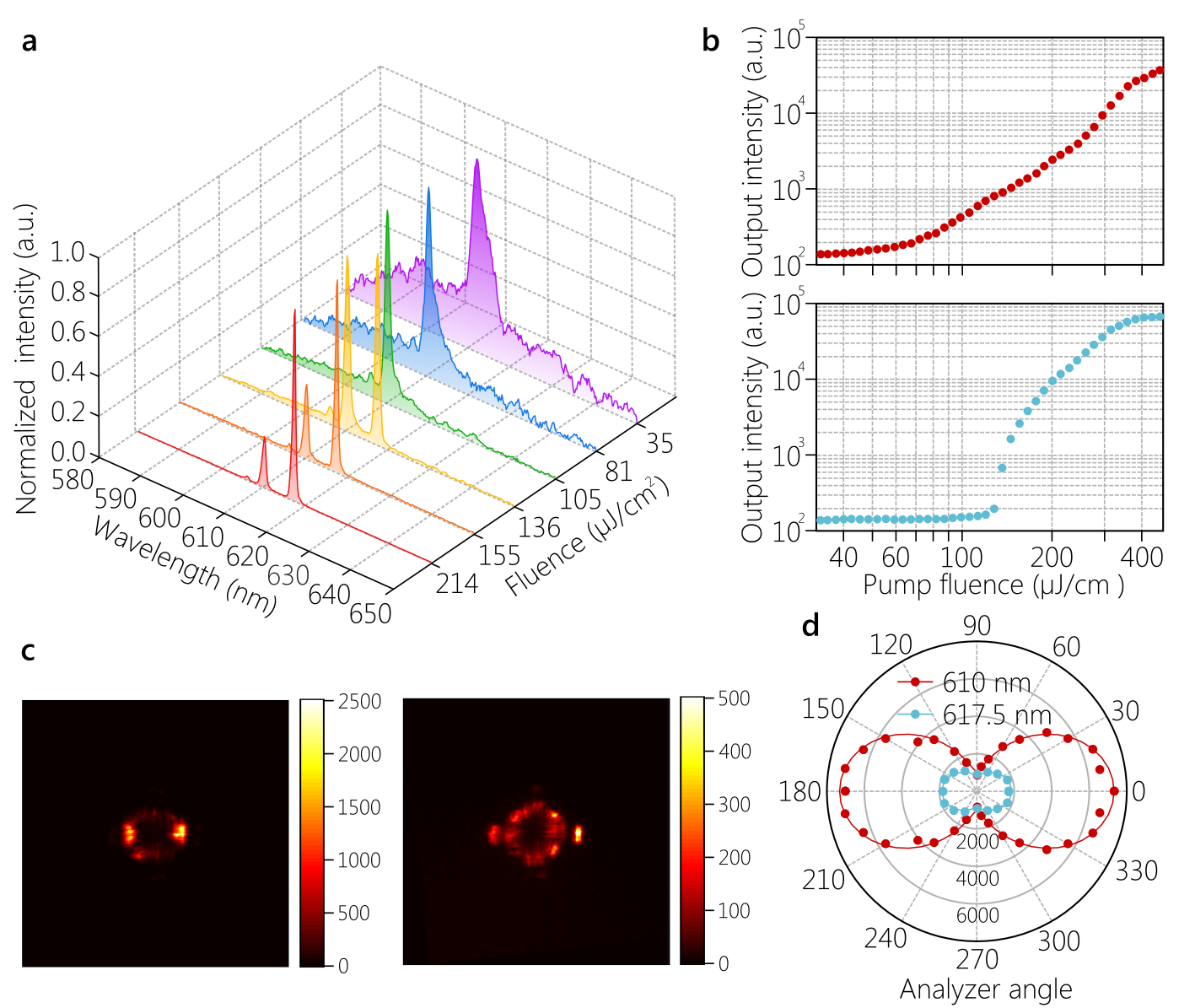}
     \caption{Dual-mode lasing action from dielectric arrays. a) The evolution of the lasing spectrum with pump fluence. b) The output intensity of the lasing peak at 610 nm (top) and 617.5 nm (bottom) as a function of pump fluence. c) Back-focal plane image of the laser emission above the threshold at an analyzer angle of 0 (left) and 90 (right). d) The intensity of the dual laser emission as a function of the polarizing analyzer angle.}
\label{lasing_fig4} 
\end{figure*}

Figure \ref{lasing_fig4} shows the lasing dynamics from an array with nanoresonators with $R_{cyl}$ = 95 nm and a $\sim$ 2 $\mu$m active cover layer of Rhodamine 101 in PVA. With increased pump intensity, one laser peak emerges at a threshold of $\sim$ 60 $\mu$J/cm$^2$ around 610 nm. At a pump fluence of 140 $\mu$J/cm$^2$, another peak appears at 617.5 nm, as seen in Fig. \ref{lasing_fig4}a. The evolution of the intensity of both laser peaks vs. the pump fluence can be seen in Fig. \ref{lasing_fig4}b. A rotating analyzer is used before feeding the collected emission to the spectrometer to study the polarization properties of the laser emission. Back-focal plane images of the laser emission at different analyzer angles are depicted in Fig. \ref{lasing_fig4}c. At 0$^o$ analyzer angle, Fig. \ref{lasing_fig4}c (left) shows that laser emission takes the shape of two spatial beams. However, an analyzer angle of 90$^o$, Fig. \ref{lasing_fig4}c (right) shows the appearance of two additional beams. The change of the measured peak values of both lasing wavelengths, 610 nm and 617 nm, is plotted in Fig. \ref{lasing_fig4}d. We can note that the laser peak at 617.5 nm shows little sensitivity to the analyzer rotation, dropping to only half of its maximum value at the analyzer angle of 90$^o$. However, the peak at 610 nm is much more sensitive to the analyzer orientation. At an analyzer angle of 90$^o$, the peak intensity of the 610 nm beam drops to $\sim$ 10\% of its maximum value (at 0$^o$ angle). From Fig. \ref{lasing_fig4}d, we can see that when the analyzer angle is set to 90$^o$, the peak amplitude of the two laser wavelengths is almost equal. From this, we conclude that each wavelength emits in a different spatial direction. The beam at 610 nm emits at 4.6$^o$ with a beamwidth of 2$^o$, and the laser peak at 617.5 nm has a spatial beam around 7.9$^o$ with a $1.6^o$ beamwidth as calculated from the back-focal plane image in Fig. \ref{lasing_fig4}c (right). Triple-mode lasing has also been observed and reported in Fig. S9.

It is worth mentioning that our choice for the unit cell for the multi-mode operation is a proof-of-concept and is simply based on the number of supported resonances. Further optimized unit cells can be investigated to provide equal peak amplitude, controllable threshold, and spectral separation. Achieving multi-mode on-chip lasing will open up a multitude of opportunities in optical computing, sensing, and telecommunications. Our design and methodology represent a step on the way to realize low threshold room-temperature nanolasers with multiple wavelengths and high directionality. Future directions will include optimization mechanisms that take full advantage of the interplay between the electric and magnetic responses in dielectric resonators alongside the rich physics of BICs to gain control on the characteristics of each lasing peak as well as their spectral separation. 

\section{Discussion} 
In summary, we design, fabricate, and experimentally demonstrate high-Q factor resonances in arrays of dielectric nanoresonators in the visible wavelength range. We utilize the proposed all-dielectric arrays to experimentally demonstrate highly-directional, low-threshold lasing action in an organic dye coating layer. To the best of our knowledge, this is the first demonstration of room temperature lasing in the visible spectrum with an all-dielectric array of nanoresonators. Our experimental studies are guided and supported by the advanced multiphysics numerical framework built on coupling carrier kinetics with a full-wave time-domain Maxwell equation solver. Using the developed framework to engineer the electric and magnetic dipole modes in dielectric resonators, we utilize the feedback from the high-Q factor structures to experimentally realize and explore multi-mode lasing. 
Our results open exciting paths for combining different physical mechanisms in a single dielectric array without intricate resonator designs.  We believe that our study could enable advanced, controllable engineering of light-matter interactions with prospective applications to the topological states engineering and quantum light generation.

\section{Methods}

\subsection{Metasurface Fabrication}
The fabrication process starts with creating the required pattern on a  silica substrate using a positive-tone resist (ZEP 520A) using electron beam lithography. The silica glass substrates (commercially available optical quality glass from PG$\&$O) are cleaned in solvent (toluene, acetone, IPA) followed by a dehydration bake at 160$^o$C for 2-5 minutes. At this stage, ZEP 520A electron-beam resist (commercially available from ZEON Chemicals) is spin-coated onto the dry substrate and cured at 180$^o$C for 3 minutes. The thickness of the resist should match the desired nanocylinder height. The spin process conditions are varied according to the spin-sheet to arrive at the desired thickness denoted as \textit{h}. Before exposing the resist to electron-beam, a very thin ($\sim$ 5 $nm$) layer of Cr metal is deposited onto the resist coated substrate using the e-beam deposition technique to avoid charging related writing discrepancies on a dielectric substrate (Fig. S6(a)). Following the deposition, the sample is immediately loaded onto the load lock of the electron beam lithography (EBL) instrument to prevent further oxidation of the Cr thin film. After electron beam exposure, the first Cr layer is etched away using Cr-16 chromium etchant solution and followed by a DI rinse. Next, the resist was developed in ZDMAC developer for 90 seconds and rinsed in IPA. Some undeveloped residual resist may remain at the bottom of the narrow cylinder holes which are then cleaned off by ashing the resist in a barrel etcher in $O_2$ plasma (at 100W, 1.2T) for 1 minute. This may lead to some reduction of the resist thickness from top-down etching as well as an increase in the diameter of the cylinders. Initial resist thickness and written diameter of the cylinders can be adjusted to account for this step. At this point, the sample should have an array of cylindrical holes in the resist (Fig. S6(b)). 

Next, the sample is loaded onto an atomic layer deposition (ALD) tool for the growth of the  TiO$_2$. ALD allows for uniform and highly conformal layer by layer growth and therefore ensures complete coverage of all the holes (Fig. S6(c)). The  TiO$_2$ thickness is decided by the maximum diameter of the cylinders (d) on the sample. Due to conformal growth from all sides thickness of $d$/2 would be enough to fill the cylinders (Fig. S6(d)). However, we keep a margin of growing an additional 10-15 $nm$. In order to ensure the stability of the polymer-based resist, the ALD operating temperature is kept much below the resist curing temperature, in this case, at a 120$^o$C. A low growth rate of 0.6 $nm$/cycle is achieved in this low-temperature process. In addition, other process parameters such as precursor and water pulse times, purge times, pressure, etc. are also optimized to arrive at a reasonably good optical quality of the grown  TiO$_2$ film. At this stage, TiO$_2$ covers not only the cylinder holes but also the top surface of the resist. To planarize the top surface and also to expose the resist layer from the top, at this step, the excess  TiO$_2$ cover is etched away using a plasma RIE technique. A combination of BCl3, Cl2, and Ar gas chemistry is used at 650W RF power, 80W bias power, and 0.8T pressure (Fig. S6(e)). 

The remaining resist matrix can then be removed in NMP (N-Methyl Pyrrolidine) heated at 120$^o$C followed by rinsing the sample in acetone and IPA. This concludes the fabrication of the high aspect ratio (Fig. S6(f)). 

\subsection{Characterization}
The fabricated metasurfaces are characterized by measuring the transmittance spectra at normal incidence for both \textit{p- and s- polarizations}. This takes place after the TiO$_2$ etching but before the removal of resist. The resist polymer has a refractive index of around 1.45 and therefore satisfies the index matching condition with the substrate. Optical characterization is performed using the J.A. Woollam V-VASE UV-VIS-NIR variable angle spectroscopic ellipsometer. The measured transmittance spectra at different $R_{cyl}$ of the resonators are given in Fig. S4(a) with solid lines showing the p-polarization and dashed lines corresponding to the s-polarization of the incident light. We can clearly observe that under normal illumination, the metasurfaces responses are polarization-independent. Additionally, the simulated transmittance as a function of $R_{cyl}$ and wavelength is shown in Fig. S4(b).  Dashed lines indicate the radii corresponding to the measurements in Fig. S4(a) with  red, blue and green indicating $R_{cyl}$ = 85 nm, $R_{cyl}$ = 90 nm and $R_{cyl}$ = 95 nm, respectively. As evident from Fig. S4, the measured transmittance of different samples is generally in good agreement with the predictions from the simulations. Deviations between experimental and simulated data are believed to be due to the fabrication imperfections as we observed non-uniformity in the radii values of the nanoresonators of the same sample. This led to spectral broadening, causing a reduction in the Q-factors as well as a shift in the location of the resonances. 

\subsection{Lasing measurements}
To study lasing properties with the metasurface feedback, we use a frequency-doubled Nd: YAG picosecond laser. The laser generates pulses with a carrier wavelength of 532 nm, a pulse width of 400 ps, and a repetition rate of 1 Hz. A schematic of the setup is shown in Fig. S7. The laser beam that pumps the metasurface is focused down to 100 $\mu$ m with a 5$\times$ objective. 
The light emission from the sample is collected with a fiber, which is then fed to a spectrometer equipped with a charge-coupled device (CCD). A notch filter centered around 532 nm is used to block the pump. All lasing measurements are done in air at room temperature. Another CCD is used to image the sample plane for sample selection and alignment and also to ensure an accurate estimation of the pump spot size.
Back focal plane measurements have been carried out with the same setup as lasing by redirecting the laser emission from the spectrometer to a CCD through a lens that is adjusted to form a back focal plane image at the camera sensor.

\subsection{FDTD simulations of lasing dynamics}
We utilize an in-home built framework of a multi-level system \cite{azzam2019exploring, azzam2018sa, azzam2018rsa} incorporated in a commercial software finite difference time domain (FDTD) solver \cite{Lum} to model nonlinear light-matter interactions \cite{yee1966fdtd, taflove2004}. The multi-level framework allows for capturing the behavior of the gain materials embedded in photonic nanostructures and have an accurate estimate of their threshold and spectral response. In this work, the gain medium (Rhodamine 101) is modeled using a four-level system, and the electromagnetic waves are treated classically with Maxwell's equations \cite{azzam2019exploring}. The four-level atomic system used is depicted in Fig. S5. More details can be found in Section 3 of SM \cite{SM}.

\section{acknowledgments}
The authors acknowledge the financial support by the U.S. Department of Energy (DOE), Office of Basic Energy Sciences (BES), Division of Materials Sciences and Engineering under Award DE-SC0017717 (experiment and fabrication) and by the DARPA/DSO Extreme Optics and Imaging (EXTREME) Program under Award HR00111720032 (modeling and simulation).

% \section{Author contributions}
% S.I.A., A. B., V.M.S., and A.V.K. conceived the research; S. I. A. performed the theoretical analysis, numerical simulations, carried out the laser experiments and wrote the original draft of the manuscript. K.C. fabricated the samples; A.L. and Y.L.K. helped with experiments; Z.J., V.M.S., A.B. and A.V.K. supervised the project. 

% \section{Competing interests}
% The authors declare no competing interests.

% \newpage
% \textbf{References}
\bibliography{refs}

\begin{thebibliography}{38}%
\makeatletter
\providecommand \@ifxundefined [1]{%
 \@ifx{#1\undefined}
}%
\providecommand \@ifnum [1]{%
 \ifnum #1\expandafter \@firstoftwo
 \else \expandafter \@secondoftwo
 \fi
}%
\providecommand \@ifx [1]{%
 \ifx #1\expandafter \@firstoftwo
 \else \expandafter \@secondoftwo
 \fi
}%
\providecommand \natexlab [1]{#1}%
\providecommand \enquote  [1]{``#1''}%
\providecommand \bibnamefont  [1]{#1}%
\providecommand \bibfnamefont [1]{#1}%
\providecommand \citenamefont [1]{#1}%
\providecommand \href@noop [0]{\@secondoftwo}%
\providecommand \href [0]{\begingroup \@sanitize@url \@href}%
\providecommand \@href[1]{\@@startlink{#1}\@@href}%
\providecommand \@@href[1]{\endgroup#1\@@endlink}%
\providecommand \@sanitize@url [0]{\catcode `\\12\catcode `\$12\catcode
  `\&12\catcode `\#12\catcode `\^12\catcode `\_12\catcode `\%12\relax}%
\providecommand \@@startlink[1]{}%
\providecommand \@@endlink[0]{}%
\providecommand \url  [0]{\begingroup\@sanitize@url \@url }%
\providecommand \@url [1]{\endgroup\@href {#1}{\urlprefix }}%
\providecommand \urlprefix  [0]{URL }%
\providecommand \Eprint [0]{\href }%
\providecommand \doibase [0]{http://dx.doi.org/}%
\providecommand \selectlanguage [0]{\@gobble}%
\providecommand \bibinfo  [0]{\@secondoftwo}%
\providecommand \bibfield  [0]{\@secondoftwo}%
\providecommand \translation [1]{[#1]}%
\providecommand \BibitemOpen [0]{}%
\providecommand \bibitemStop [0]{}%
\providecommand \bibitemNoStop [0]{.\EOS\space}%
\providecommand \EOS [0]{\spacefactor3000\relax}%
\providecommand \BibitemShut  [1]{\csname bibitem#1\endcsname}%
\let\auto@bib@innerbib\@empty
%</preamble>
\bibitem [{\citenamefont {von Neumann}\ and\ \citenamefont
  {Wigner}(1929)}]{von1929some}%
  \BibitemOpen
  \bibfield  {author} {\bibinfo {author} {\bibfnamefont {J.}~\bibnamefont {von
  Neumann}}\ and\ \bibinfo {author} {\bibfnamefont {E.}~\bibnamefont
  {Wigner}},\ }\href@noop {} {\bibfield  {journal} {\bibinfo  {journal} {Phys.
  Z}\ }\textbf {\bibinfo {volume} {465}} (\bibinfo {year} {1929})}\BibitemShut
  {NoStop}%
\bibitem [{\citenamefont {Lee}\ \emph {et~al.}(2012)\citenamefont {Lee},
  \citenamefont {Zhen}, \citenamefont {Chua}, \citenamefont {Qiu},
  \citenamefont {Joannopoulos}, \citenamefont {Solja{\v{c}}i{\'c}},\ and\
  \citenamefont {Shapira}}]{lee2012observation}%
  \BibitemOpen
  \bibfield  {author} {\bibinfo {author} {\bibfnamefont {J.}~\bibnamefont
  {Lee}}, \bibinfo {author} {\bibfnamefont {B.}~\bibnamefont {Zhen}}, \bibinfo
  {author} {\bibfnamefont {S.-L.}\ \bibnamefont {Chua}}, \bibinfo {author}
  {\bibfnamefont {W.}~\bibnamefont {Qiu}}, \bibinfo {author} {\bibfnamefont
  {J.~D.}\ \bibnamefont {Joannopoulos}}, \bibinfo {author} {\bibfnamefont
  {M.}~\bibnamefont {Solja{\v{c}}i{\'c}}}, \ and\ \bibinfo {author}
  {\bibfnamefont {O.}~\bibnamefont {Shapira}},\ }\href@noop {} {\bibfield
  {journal} {\bibinfo  {journal} {Phys. Rev. Lett.}\ }\textbf {\bibinfo
  {volume} {109}},\ \bibinfo {pages} {067401} (\bibinfo {year}
  {2012})}\BibitemShut {NoStop}%
\bibitem [{\citenamefont {Plotnik}\ \emph {et~al.}(2011)\citenamefont
  {Plotnik}, \citenamefont {Peleg}, \citenamefont {Dreisow}, \citenamefont
  {Heinrich}, \citenamefont {Nolte}, \citenamefont {Szameit},\ and\
  \citenamefont {Segev}}]{plotnik2011experimental}%
  \BibitemOpen
  \bibfield  {author} {\bibinfo {author} {\bibfnamefont {Y.}~\bibnamefont
  {Plotnik}}, \bibinfo {author} {\bibfnamefont {O.}~\bibnamefont {Peleg}},
  \bibinfo {author} {\bibfnamefont {F.}~\bibnamefont {Dreisow}}, \bibinfo
  {author} {\bibfnamefont {M.}~\bibnamefont {Heinrich}}, \bibinfo {author}
  {\bibfnamefont {S.}~\bibnamefont {Nolte}}, \bibinfo {author} {\bibfnamefont
  {A.}~\bibnamefont {Szameit}}, \ and\ \bibinfo {author} {\bibfnamefont
  {M.}~\bibnamefont {Segev}},\ }\href@noop {} {\bibfield  {journal} {\bibinfo
  {journal} {Phys. Rev. Lett.}\ }\textbf {\bibinfo {volume} {107}},\ \bibinfo
  {pages} {183901} (\bibinfo {year} {2011})}\BibitemShut {NoStop}%
\bibitem [{\citenamefont {Moiseyev}(2009)}]{moiseyev2009suppression}%
  \BibitemOpen
  \bibfield  {author} {\bibinfo {author} {\bibfnamefont {N.}~\bibnamefont
  {Moiseyev}},\ }\href@noop {} {\bibfield  {journal} {\bibinfo  {journal}
  {Phys. Rev. Lett.}\ }\textbf {\bibinfo {volume} {102}},\ \bibinfo {pages}
  {167404} (\bibinfo {year} {2009})}\BibitemShut {NoStop}%
\bibitem [{\citenamefont {Friedrich}\ and\ \citenamefont
  {Wintgen}(1985)}]{friedrich1985interfering}%
  \BibitemOpen
  \bibfield  {author} {\bibinfo {author} {\bibfnamefont {H.}~\bibnamefont
  {Friedrich}}\ and\ \bibinfo {author} {\bibfnamefont {D.}~\bibnamefont
  {Wintgen}},\ }\href@noop {} {\bibfield  {journal} {\bibinfo  {journal} {Phys.
  Rev. A}\ }\textbf {\bibinfo {volume} {32}},\ \bibinfo {pages} {3231}
  (\bibinfo {year} {1985})}\BibitemShut {NoStop}%
\bibitem [{\citenamefont {Hsu}\ \emph {et~al.}(2016)\citenamefont {Hsu},
  \citenamefont {Zhen}, \citenamefont {Stone}, \citenamefont {Joannopoulos},\
  and\ \citenamefont {Solja{\v{c}}i{\'c}}}]{hsu2016bound}%
  \BibitemOpen
  \bibfield  {author} {\bibinfo {author} {\bibfnamefont {C.~W.}\ \bibnamefont
  {Hsu}}, \bibinfo {author} {\bibfnamefont {B.}~\bibnamefont {Zhen}}, \bibinfo
  {author} {\bibfnamefont {A.~D.}\ \bibnamefont {Stone}}, \bibinfo {author}
  {\bibfnamefont {J.~D.}\ \bibnamefont {Joannopoulos}}, \ and\ \bibinfo
  {author} {\bibfnamefont {M.}~\bibnamefont {Solja{\v{c}}i{\'c}}},\ }\href@noop
  {} {\bibfield  {journal} {\bibinfo  {journal} {Nature Reviews Materials}\
  }\textbf {\bibinfo {volume} {1}},\ \bibinfo {pages} {16048} (\bibinfo {year}
  {2016})}\BibitemShut {NoStop}%
\bibitem [{\citenamefont {Doeleman}\ \emph {et~al.}(2018)\citenamefont
  {Doeleman}, \citenamefont {Monticone}, \citenamefont {den Hollander},
  \citenamefont {Al{\`u}},\ and\ \citenamefont
  {Koenderink}}]{alu2018experimental}%
  \BibitemOpen
  \bibfield  {author} {\bibinfo {author} {\bibfnamefont {H.~M.}\ \bibnamefont
  {Doeleman}}, \bibinfo {author} {\bibfnamefont {F.}~\bibnamefont {Monticone}},
  \bibinfo {author} {\bibfnamefont {W.}~\bibnamefont {den Hollander}}, \bibinfo
  {author} {\bibfnamefont {A.}~\bibnamefont {Al{\`u}}}, \ and\ \bibinfo
  {author} {\bibfnamefont {A.~F.}\ \bibnamefont {Koenderink}},\ }\href
  {\doibase 10.1038/s41566-018-0177-5} {\bibfield  {journal} {\bibinfo
  {journal} {Nat. Photonics}\ }\textbf {\bibinfo {volume} {12}},\ \bibinfo
  {pages} {397} (\bibinfo {year} {2018})}\BibitemShut {NoStop}%
\bibitem [{\citenamefont {Hsu}\ \emph {et~al.}(2013)\citenamefont {Hsu},
  \citenamefont {Zhen}, \citenamefont {Lee}, \citenamefont {Chua},
  \citenamefont {Johnson}, \citenamefont {Joannopoulos},\ and\ \citenamefont
  {Solja{\v c}i{\'c}}}]{hsu2013observation}%
  \BibitemOpen
  \bibfield  {author} {\bibinfo {author} {\bibfnamefont {C.~W.}\ \bibnamefont
  {Hsu}}, \bibinfo {author} {\bibfnamefont {B.}~\bibnamefont {Zhen}}, \bibinfo
  {author} {\bibfnamefont {J.}~\bibnamefont {Lee}}, \bibinfo {author}
  {\bibfnamefont {S.-L.}\ \bibnamefont {Chua}}, \bibinfo {author}
  {\bibfnamefont {S.~G.}\ \bibnamefont {Johnson}}, \bibinfo {author}
  {\bibfnamefont {J.~D.}\ \bibnamefont {Joannopoulos}}, \ and\ \bibinfo
  {author} {\bibfnamefont {M.}~\bibnamefont {Solja{\v c}i{\'c}}},\ }\href@noop
  {} {\bibfield  {journal} {\bibinfo  {journal} {Nature}\ }\textbf {\bibinfo
  {volume} {499}},\ \bibinfo {pages} {188 } (\bibinfo {year}
  {2013})}\BibitemShut {NoStop}%
\bibitem [{\citenamefont {Sadrieva}\ \emph {et~al.}(2017)\citenamefont
  {Sadrieva}, \citenamefont {Sinev}, \citenamefont {Koshelev}, \citenamefont
  {Samusev}, \citenamefont {Iorsh}, \citenamefont {Takayama}, \citenamefont
  {Malureanu}, \citenamefont {Bogdanov},\ and\ \citenamefont
  {Lavrinenko}}]{sadrieva2017transition}%
  \BibitemOpen
  \bibfield  {author} {\bibinfo {author} {\bibfnamefont {Z.~F.}\ \bibnamefont
  {Sadrieva}}, \bibinfo {author} {\bibfnamefont {I.~S.}\ \bibnamefont {Sinev}},
  \bibinfo {author} {\bibfnamefont {K.~L.}\ \bibnamefont {Koshelev}}, \bibinfo
  {author} {\bibfnamefont {A.}~\bibnamefont {Samusev}}, \bibinfo {author}
  {\bibfnamefont {I.~V.}\ \bibnamefont {Iorsh}}, \bibinfo {author}
  {\bibfnamefont {O.}~\bibnamefont {Takayama}}, \bibinfo {author}
  {\bibfnamefont {R.}~\bibnamefont {Malureanu}}, \bibinfo {author}
  {\bibfnamefont {A.~A.}\ \bibnamefont {Bogdanov}}, \ and\ \bibinfo {author}
  {\bibfnamefont {A.~V.}\ \bibnamefont {Lavrinenko}},\ }\href@noop {}
  {\bibfield  {journal} {\bibinfo  {journal} {ACS Photonics}\ }\textbf
  {\bibinfo {volume} {4}},\ \bibinfo {pages} {723} (\bibinfo {year}
  {2017})}\BibitemShut {NoStop}%
\bibitem [{\citenamefont {Azzam}\ \emph {et~al.}(2018)\citenamefont {Azzam},
  \citenamefont {Shalaev}, \citenamefont {Boltasseva},\ and\ \citenamefont
  {Kildishev}}]{azzam2018formation}%
  \BibitemOpen
  \bibfield  {author} {\bibinfo {author} {\bibfnamefont {S.~I.}\ \bibnamefont
  {Azzam}}, \bibinfo {author} {\bibfnamefont {V.~M.}\ \bibnamefont {Shalaev}},
  \bibinfo {author} {\bibfnamefont {A.}~\bibnamefont {Boltasseva}}, \ and\
  \bibinfo {author} {\bibfnamefont {A.~V.}\ \bibnamefont {Kildishev}},\
  }\href@noop {} {\bibfield  {journal} {\bibinfo  {journal} {Physical review
  letters}\ }\textbf {\bibinfo {volume} {121}},\ \bibinfo {pages} {253901}
  (\bibinfo {year} {2018})}\BibitemShut {NoStop}%
\bibitem [{\citenamefont {Fonda}(1961)}]{fonda1961resonance}%
  \BibitemOpen
  \bibfield  {author} {\bibinfo {author} {\bibfnamefont {L.}~\bibnamefont
  {Fonda}},\ }\href@noop {} {\bibfield  {journal} {\bibinfo  {journal} {Annals
  of Physics}\ }\textbf {\bibinfo {volume} {12}},\ \bibinfo {pages} {476}
  (\bibinfo {year} {1961})}\BibitemShut {NoStop}%
\bibitem [{\citenamefont {Fonda}(1963)}]{fonda1963bound}%
  \BibitemOpen
  \bibfield  {author} {\bibinfo {author} {\bibfnamefont {L.}~\bibnamefont
  {Fonda}},\ }\href@noop {} {\bibfield  {journal} {\bibinfo  {journal} {Annals
  of Physics}\ }\textbf {\bibinfo {volume} {22}},\ \bibinfo {pages} {123}
  (\bibinfo {year} {1963})}\BibitemShut {NoStop}%
\bibitem [{\citenamefont {Marinica}\ \emph {et~al.}(2008)\citenamefont
  {Marinica}, \citenamefont {Borisov},\ and\ \citenamefont
  {Shabanov}}]{marinica2008bound}%
  \BibitemOpen
  \bibfield  {author} {\bibinfo {author} {\bibfnamefont {D.~C.}\ \bibnamefont
  {Marinica}}, \bibinfo {author} {\bibfnamefont {A.~G.}\ \bibnamefont
  {Borisov}}, \ and\ \bibinfo {author} {\bibfnamefont {S.~V.}\ \bibnamefont
  {Shabanov}},\ }\href@noop {} {\bibfield  {journal} {\bibinfo  {journal}
  {Phys. Rev. Lett.}\ }\textbf {\bibinfo {volume} {100}},\ \bibinfo {pages}
  {183902} (\bibinfo {year} {2008})}\BibitemShut {NoStop}%
\bibitem [{\citenamefont {Rybin}\ \emph {et~al.}(2017)\citenamefont {Rybin},
  \citenamefont {Koshelev}, \citenamefont {Sadrieva}, \citenamefont {Samusev},
  \citenamefont {Bogdanov}, \citenamefont {Limonov},\ and\ \citenamefont
  {Kivshar}}]{kivshar2017high}%
  \BibitemOpen
  \bibfield  {author} {\bibinfo {author} {\bibfnamefont {M.~V.}\ \bibnamefont
  {Rybin}}, \bibinfo {author} {\bibfnamefont {K.~L.}\ \bibnamefont {Koshelev}},
  \bibinfo {author} {\bibfnamefont {Z.~F.}\ \bibnamefont {Sadrieva}}, \bibinfo
  {author} {\bibfnamefont {K.~B.}\ \bibnamefont {Samusev}}, \bibinfo {author}
  {\bibfnamefont {A.~A.}\ \bibnamefont {Bogdanov}}, \bibinfo {author}
  {\bibfnamefont {M.~F.}\ \bibnamefont {Limonov}}, \ and\ \bibinfo {author}
  {\bibfnamefont {Y.~S.}\ \bibnamefont {Kivshar}},\ }\href@noop {} {\bibfield
  {journal} {\bibinfo  {journal} {Phys. Rev. Lett.}\ }\textbf {\bibinfo
  {volume} {119}},\ \bibinfo {pages} {243901} (\bibinfo {year}
  {2017})}\BibitemShut {NoStop}%
\bibitem [{\citenamefont {Krasnok}\ and\ \citenamefont
  {Al{\'u}}(2018)}]{alu2018embedded}%
  \BibitemOpen
  \bibfield  {author} {\bibinfo {author} {\bibfnamefont {A.}~\bibnamefont
  {Krasnok}}\ and\ \bibinfo {author} {\bibfnamefont {A.}~\bibnamefont
  {Al{\'u}}},\ }\href@noop {} {\bibfield  {journal} {\bibinfo  {journal} {J.
  Opt.}\ }\textbf {\bibinfo {volume} {20}},\ \bibinfo {pages} {064002}
  (\bibinfo {year} {2018})}\BibitemShut {NoStop}%
\bibitem [{\citenamefont {Bulgakov}\ and\ \citenamefont
  {Sadreev}(2008)}]{Sadreev2008bound}%
  \BibitemOpen
  \bibfield  {author} {\bibinfo {author} {\bibfnamefont {E.~N.}\ \bibnamefont
  {Bulgakov}}\ and\ \bibinfo {author} {\bibfnamefont {A.~F.}\ \bibnamefont
  {Sadreev}},\ }\href@noop {} {\bibfield  {journal} {\bibinfo  {journal} {Phy.
  Rev. B}\ }\textbf {\bibinfo {volume} {78}},\ \bibinfo {pages} {075105}
  (\bibinfo {year} {2008})}\BibitemShut {NoStop}%
\bibitem [{\citenamefont {Yu}\ \emph {et~al.}(2019)\citenamefont {Yu},
  \citenamefont {Xi}, \citenamefont {Ma}, \citenamefont {Tsang}, \citenamefont
  {Zou},\ and\ \citenamefont {Sun}}]{yu2019photonic}%
  \BibitemOpen
  \bibfield  {author} {\bibinfo {author} {\bibfnamefont {Z.}~\bibnamefont
  {Yu}}, \bibinfo {author} {\bibfnamefont {X.}~\bibnamefont {Xi}}, \bibinfo
  {author} {\bibfnamefont {J.}~\bibnamefont {Ma}}, \bibinfo {author}
  {\bibfnamefont {H.~K.}\ \bibnamefont {Tsang}}, \bibinfo {author}
  {\bibfnamefont {C.-L.}\ \bibnamefont {Zou}}, \ and\ \bibinfo {author}
  {\bibfnamefont {X.}~\bibnamefont {Sun}},\ }\href@noop {} {\bibfield
  {journal} {\bibinfo  {journal} {Optica}\ }\textbf {\bibinfo {volume} {6}},\
  \bibinfo {pages} {1342} (\bibinfo {year} {2019})}\BibitemShut {NoStop}%
\bibitem [{\citenamefont {Koshelev}\ \emph {et~al.}(2020)\citenamefont
  {Koshelev}, \citenamefont {Kruk}, \citenamefont {Melik-Gaykazyan},
  \citenamefont {Choi}, \citenamefont {Bogdanov}, \citenamefont {Park},\ and\
  \citenamefont {Kivshar}}]{koshelev2020subwavelength}%
  \BibitemOpen
  \bibfield  {author} {\bibinfo {author} {\bibfnamefont {K.}~\bibnamefont
  {Koshelev}}, \bibinfo {author} {\bibfnamefont {S.}~\bibnamefont {Kruk}},
  \bibinfo {author} {\bibfnamefont {E.}~\bibnamefont {Melik-Gaykazyan}},
  \bibinfo {author} {\bibfnamefont {J.-H.}\ \bibnamefont {Choi}}, \bibinfo
  {author} {\bibfnamefont {A.}~\bibnamefont {Bogdanov}}, \bibinfo {author}
  {\bibfnamefont {H.-G.}\ \bibnamefont {Park}}, \ and\ \bibinfo {author}
  {\bibfnamefont {Y.}~\bibnamefont {Kivshar}},\ }\href@noop {} {\bibfield
  {journal} {\bibinfo  {journal} {Science}\ }\textbf {\bibinfo {volume}
  {367}},\ \bibinfo {pages} {288} (\bibinfo {year} {2020})}\BibitemShut
  {NoStop}%
\bibitem [{\citenamefont {Leitis}\ \emph {et~al.}(2019)\citenamefont {Leitis},
  \citenamefont {Tittl}, \citenamefont {Liu}, \citenamefont {Lee},
  \citenamefont {Gu}, \citenamefont {Kivshar},\ and\ \citenamefont
  {Altug}}]{leitis2019angle}%
  \BibitemOpen
  \bibfield  {author} {\bibinfo {author} {\bibfnamefont {A.}~\bibnamefont
  {Leitis}}, \bibinfo {author} {\bibfnamefont {A.}~\bibnamefont {Tittl}},
  \bibinfo {author} {\bibfnamefont {M.}~\bibnamefont {Liu}}, \bibinfo {author}
  {\bibfnamefont {B.~H.}\ \bibnamefont {Lee}}, \bibinfo {author} {\bibfnamefont
  {M.~B.}\ \bibnamefont {Gu}}, \bibinfo {author} {\bibfnamefont {Y.~S.}\
  \bibnamefont {Kivshar}}, \ and\ \bibinfo {author} {\bibfnamefont
  {H.}~\bibnamefont {Altug}},\ }\href@noop {} {\bibfield  {journal} {\bibinfo
  {journal} {Science advances}\ }\textbf {\bibinfo {volume} {5}},\ \bibinfo
  {pages} {eaaw2871} (\bibinfo {year} {2019})}\BibitemShut {NoStop}%
\bibitem [{\citenamefont {Kodigala}\ \emph {et~al.}(2017)\citenamefont
  {Kodigala}, \citenamefont {Lepetit}, \citenamefont {Gu}, \citenamefont
  {Bahari}, \citenamefont {Fainman},\ and\ \citenamefont
  {Kant{\'e}}}]{kante2017lasing}%
  \BibitemOpen
  \bibfield  {author} {\bibinfo {author} {\bibfnamefont {A.}~\bibnamefont
  {Kodigala}}, \bibinfo {author} {\bibfnamefont {T.}~\bibnamefont {Lepetit}},
  \bibinfo {author} {\bibfnamefont {Q.}~\bibnamefont {Gu}}, \bibinfo {author}
  {\bibfnamefont {B.}~\bibnamefont {Bahari}}, \bibinfo {author} {\bibfnamefont
  {Y.}~\bibnamefont {Fainman}}, \ and\ \bibinfo {author} {\bibfnamefont
  {B.}~\bibnamefont {Kant{\'e}}},\ }\href@noop {} {\bibfield  {journal}
  {\bibinfo  {journal} {Nature}\ }\textbf {\bibinfo {volume} {541}},\ \bibinfo
  {pages} {196} (\bibinfo {year} {2017})}\BibitemShut {NoStop}%
\bibitem [{\citenamefont {Ha}\ \emph {et~al.}(2018)\citenamefont {Ha},
  \citenamefont {Fu}, \citenamefont {Emani}, \citenamefont {Pan}, \citenamefont
  {Bakker}, \citenamefont {Paniagua-Dom{\'\i}nguez},\ and\ \citenamefont
  {Kuznetsov}}]{Arseniy2018}%
  \BibitemOpen
  \bibfield  {author} {\bibinfo {author} {\bibfnamefont {S.~T.}\ \bibnamefont
  {Ha}}, \bibinfo {author} {\bibfnamefont {Y.~H.}\ \bibnamefont {Fu}}, \bibinfo
  {author} {\bibfnamefont {N.~K.}\ \bibnamefont {Emani}}, \bibinfo {author}
  {\bibfnamefont {Z.}~\bibnamefont {Pan}}, \bibinfo {author} {\bibfnamefont
  {R.~M.}\ \bibnamefont {Bakker}}, \bibinfo {author} {\bibfnamefont
  {R.}~\bibnamefont {Paniagua-Dom{\'\i}nguez}}, \ and\ \bibinfo {author}
  {\bibfnamefont {A.~I.}\ \bibnamefont {Kuznetsov}},\ }\href@noop {} {\bibfield
   {journal} {\bibinfo  {journal} {Nat. Nanotech.}\ ,\ \bibinfo {pages} {1}}
  (\bibinfo {year} {2018})}\BibitemShut {NoStop}%
\bibitem [{\citenamefont {Huang}\ \emph {et~al.}(2020)\citenamefont {Huang},
  \citenamefont {Zhang}, \citenamefont {Xiao}, \citenamefont {Wang},
  \citenamefont {Fan}, \citenamefont {Liu}, \citenamefont {Zhang},
  \citenamefont {Qu}, \citenamefont {Ji}, \citenamefont {Han} \emph
  {et~al.}}]{huang2020ultrafast}%
  \BibitemOpen
  \bibfield  {author} {\bibinfo {author} {\bibfnamefont {C.}~\bibnamefont
  {Huang}}, \bibinfo {author} {\bibfnamefont {C.}~\bibnamefont {Zhang}},
  \bibinfo {author} {\bibfnamefont {S.}~\bibnamefont {Xiao}}, \bibinfo {author}
  {\bibfnamefont {Y.}~\bibnamefont {Wang}}, \bibinfo {author} {\bibfnamefont
  {Y.}~\bibnamefont {Fan}}, \bibinfo {author} {\bibfnamefont {Y.}~\bibnamefont
  {Liu}}, \bibinfo {author} {\bibfnamefont {N.}~\bibnamefont {Zhang}}, \bibinfo
  {author} {\bibfnamefont {G.}~\bibnamefont {Qu}}, \bibinfo {author}
  {\bibfnamefont {H.}~\bibnamefont {Ji}}, \bibinfo {author} {\bibfnamefont
  {J.}~\bibnamefont {Han}},  \emph {et~al.},\ }\href@noop {} {\bibfield
  {journal} {\bibinfo  {journal} {Science}\ }\textbf {\bibinfo {volume}
  {367}},\ \bibinfo {pages} {1018} (\bibinfo {year} {2020})}\BibitemShut
  {NoStop}%
\bibitem [{\citenamefont {Kuznetsov}\ \emph {et~al.}(2016)\citenamefont
  {Kuznetsov}, \citenamefont {Miroshnichenko}, \citenamefont {Brongersma},
  \citenamefont {Kivshar},\ and\ \citenamefont
  {Luk’yanchuk}}]{kuznetsov2016optically}%
  \BibitemOpen
  \bibfield  {author} {\bibinfo {author} {\bibfnamefont {A.~I.}\ \bibnamefont
  {Kuznetsov}}, \bibinfo {author} {\bibfnamefont {A.~E.}\ \bibnamefont
  {Miroshnichenko}}, \bibinfo {author} {\bibfnamefont {M.~L.}\ \bibnamefont
  {Brongersma}}, \bibinfo {author} {\bibfnamefont {Y.~S.}\ \bibnamefont
  {Kivshar}}, \ and\ \bibinfo {author} {\bibfnamefont {B.}~\bibnamefont
  {Luk’yanchuk}},\ }\href@noop {} {\bibfield  {journal} {\bibinfo  {journal}
  {Science}\ }\textbf {\bibinfo {volume} {354}},\ \bibinfo {pages} {aag2472}
  (\bibinfo {year} {2016})}\BibitemShut {NoStop}%
\bibitem [{\citenamefont {Limonov}\ \emph {et~al.}(2017)\citenamefont
  {Limonov}, \citenamefont {Rybin}, \citenamefont {Poddubny},\ and\
  \citenamefont {Kivshar}}]{kivshar2017nature_review}%
  \BibitemOpen
  \bibfield  {author} {\bibinfo {author} {\bibfnamefont {M.~F.}\ \bibnamefont
  {Limonov}}, \bibinfo {author} {\bibfnamefont {M.~V.}\ \bibnamefont {Rybin}},
  \bibinfo {author} {\bibfnamefont {A.~N.}\ \bibnamefont {Poddubny}}, \ and\
  \bibinfo {author} {\bibfnamefont {Y.~S.}\ \bibnamefont {Kivshar}},\
  }\href@noop {} {\bibfield  {journal} {\bibinfo  {journal} {Nat. Photonics}\
  }\textbf {\bibinfo {volume} {11}},\ \bibinfo {pages} {543} (\bibinfo {year}
  {2017})}\BibitemShut {NoStop}%
\bibitem [{\citenamefont {Van~de Groep}\ and\ \citenamefont
  {Polman}(2013)}]{van2013designing}%
  \BibitemOpen
  \bibfield  {author} {\bibinfo {author} {\bibfnamefont {J.}~\bibnamefont
  {Van~de Groep}}\ and\ \bibinfo {author} {\bibfnamefont {A.}~\bibnamefont
  {Polman}},\ }\href@noop {} {\bibfield  {journal} {\bibinfo  {journal} {Optics
  express}\ }\textbf {\bibinfo {volume} {21}},\ \bibinfo {pages} {26285}
  (\bibinfo {year} {2013})}\BibitemShut {NoStop}%
\bibitem [{\citenamefont {Evlyukhin}\ \emph {et~al.}(2011)\citenamefont
  {Evlyukhin}, \citenamefont {Reinhardt},\ and\ \citenamefont
  {Chichkov}}]{evlyukhin2011multipole}%
  \BibitemOpen
  \bibfield  {author} {\bibinfo {author} {\bibfnamefont {A.~B.}\ \bibnamefont
  {Evlyukhin}}, \bibinfo {author} {\bibfnamefont {C.}~\bibnamefont
  {Reinhardt}}, \ and\ \bibinfo {author} {\bibfnamefont {B.~N.}\ \bibnamefont
  {Chichkov}},\ }\href@noop {} {\bibfield  {journal} {\bibinfo  {journal}
  {Physical Review B}\ }\textbf {\bibinfo {volume} {84}},\ \bibinfo {pages}
  {235429} (\bibinfo {year} {2011})}\BibitemShut {NoStop}%
\bibitem [{\citenamefont {Babicheva}\ and\ \citenamefont
  {Moloney}(2018)}]{babicheva2018lattice}%
  \BibitemOpen
  \bibfield  {author} {\bibinfo {author} {\bibfnamefont {V.~E.}\ \bibnamefont
  {Babicheva}}\ and\ \bibinfo {author} {\bibfnamefont {J.~V.}\ \bibnamefont
  {Moloney}},\ }\href@noop {} {\bibfield  {journal} {\bibinfo  {journal}
  {Nanophotonics}\ }\textbf {\bibinfo {volume} {7}},\ \bibinfo {pages} {1663}
  (\bibinfo {year} {2018})}\BibitemShut {NoStop}%
\bibitem [{\citenamefont {Li}\ \emph {et~al.}(2018)\citenamefont {Li},
  \citenamefont {Verellen},\ and\ \citenamefont
  {Van~Dorpe}}]{li2018engineering}%
  \BibitemOpen
  \bibfield  {author} {\bibinfo {author} {\bibfnamefont {J.}~\bibnamefont
  {Li}}, \bibinfo {author} {\bibfnamefont {N.}~\bibnamefont {Verellen}}, \ and\
  \bibinfo {author} {\bibfnamefont {P.}~\bibnamefont {Van~Dorpe}},\ }\href@noop
  {} {\bibfield  {journal} {\bibinfo  {journal} {Journal of Applied Physics}\
  }\textbf {\bibinfo {volume} {123}},\ \bibinfo {pages} {083101} (\bibinfo
  {year} {2018})}\BibitemShut {NoStop}%
\bibitem [{SM()}]{SM}%
  \BibitemOpen
  \href@noop {} {}\bibinfo {note} {See supplementary materials.}\BibitemShut
  {Stop}%
\bibitem [{\citenamefont {Azzam}\ \emph {et~al.}(2019)\citenamefont {Azzam},
  \citenamefont {Fang}, \citenamefont {Liu}, \citenamefont {Wang},
  \citenamefont {Arnold}, \citenamefont {Klar}, \citenamefont {Prokopeva},
  \citenamefont {Meng}, \citenamefont {Shalaev},\ and\ \citenamefont
  {Kildishev}}]{azzam2019exploring}%
  \BibitemOpen
  \bibfield  {author} {\bibinfo {author} {\bibfnamefont {S.~I.}\ \bibnamefont
  {Azzam}}, \bibinfo {author} {\bibfnamefont {J.}~\bibnamefont {Fang}},
  \bibinfo {author} {\bibfnamefont {J.}~\bibnamefont {Liu}}, \bibinfo {author}
  {\bibfnamefont {Z.}~\bibnamefont {Wang}}, \bibinfo {author} {\bibfnamefont
  {N.}~\bibnamefont {Arnold}}, \bibinfo {author} {\bibfnamefont {T.~A.}\
  \bibnamefont {Klar}}, \bibinfo {author} {\bibfnamefont {L.~J.}\ \bibnamefont
  {Prokopeva}}, \bibinfo {author} {\bibfnamefont {X.}~\bibnamefont {Meng}},
  \bibinfo {author} {\bibfnamefont {V.~M.}\ \bibnamefont {Shalaev}}, \ and\
  \bibinfo {author} {\bibfnamefont {A.~V.}\ \bibnamefont {Kildishev}},\
  }\href@noop {} {\bibfield  {journal} {\bibinfo  {journal} {Laser \& Photonics
  Reviews}\ }\textbf {\bibinfo {volume} {13}},\ \bibinfo {pages} {1800071}
  (\bibinfo {year} {2019})}\BibitemShut {NoStop}%
\bibitem [{\citenamefont {Yang}\ \emph {et~al.}(2014)\citenamefont {Yang},
  \citenamefont {Kravchenko}, \citenamefont {Briggs},\ and\ \citenamefont
  {Valentine}}]{yang2014eit}%
  \BibitemOpen
  \bibfield  {author} {\bibinfo {author} {\bibfnamefont {Y.}~\bibnamefont
  {Yang}}, \bibinfo {author} {\bibfnamefont {I.~I.}\ \bibnamefont
  {Kravchenko}}, \bibinfo {author} {\bibfnamefont {D.~P.}\ \bibnamefont
  {Briggs}}, \ and\ \bibinfo {author} {\bibfnamefont {J.}~\bibnamefont
  {Valentine}},\ }\href@noop {} {\bibfield  {journal} {\bibinfo  {journal}
  {Nature communications}\ }\textbf {\bibinfo {volume} {5}},\ \bibinfo {pages}
  {1} (\bibinfo {year} {2014})}\BibitemShut {NoStop}%
\bibitem [{\citenamefont {Campione}\ \emph {et~al.}(2016)\citenamefont
  {Campione}, \citenamefont {Liu}, \citenamefont {Basilio}, \citenamefont
  {Warne}, \citenamefont {Langston}, \citenamefont {Luk}, \citenamefont
  {Wendt}, \citenamefont {Reno}, \citenamefont {Keeler}, \citenamefont {Brener}
  \emph {et~al.}}]{campione2016broken}%
  \BibitemOpen
  \bibfield  {author} {\bibinfo {author} {\bibfnamefont {S.}~\bibnamefont
  {Campione}}, \bibinfo {author} {\bibfnamefont {S.}~\bibnamefont {Liu}},
  \bibinfo {author} {\bibfnamefont {L.~I.}\ \bibnamefont {Basilio}}, \bibinfo
  {author} {\bibfnamefont {L.~K.}\ \bibnamefont {Warne}}, \bibinfo {author}
  {\bibfnamefont {W.~L.}\ \bibnamefont {Langston}}, \bibinfo {author}
  {\bibfnamefont {T.~S.}\ \bibnamefont {Luk}}, \bibinfo {author} {\bibfnamefont
  {J.~R.}\ \bibnamefont {Wendt}}, \bibinfo {author} {\bibfnamefont {J.~L.}\
  \bibnamefont {Reno}}, \bibinfo {author} {\bibfnamefont {G.~A.}\ \bibnamefont
  {Keeler}}, \bibinfo {author} {\bibfnamefont {I.}~\bibnamefont {Brener}},
  \emph {et~al.},\ }\href@noop {} {\bibfield  {journal} {\bibinfo  {journal}
  {Acs Photonics}\ }\textbf {\bibinfo {volume} {3}},\ \bibinfo {pages} {2362}
  (\bibinfo {year} {2016})}\BibitemShut {NoStop}%
\bibitem [{\citenamefont {Wang}\ \emph {et~al.}(2017)\citenamefont {Wang},
  \citenamefont {Yang}, \citenamefont {Wang}, \citenamefont {Hua},
  \citenamefont {Schaller}, \citenamefont {Schatz},\ and\ \citenamefont
  {Odom}}]{wang2017band}%
  \BibitemOpen
  \bibfield  {author} {\bibinfo {author} {\bibfnamefont {D.}~\bibnamefont
  {Wang}}, \bibinfo {author} {\bibfnamefont {A.}~\bibnamefont {Yang}}, \bibinfo
  {author} {\bibfnamefont {W.}~\bibnamefont {Wang}}, \bibinfo {author}
  {\bibfnamefont {Y.}~\bibnamefont {Hua}}, \bibinfo {author} {\bibfnamefont
  {R.~D.}\ \bibnamefont {Schaller}}, \bibinfo {author} {\bibfnamefont {G.~C.}\
  \bibnamefont {Schatz}}, \ and\ \bibinfo {author} {\bibfnamefont {T.~W.}\
  \bibnamefont {Odom}},\ }\href@noop {} {\bibfield  {journal} {\bibinfo
  {journal} {Nature nanotechnology}\ }\textbf {\bibinfo {volume} {12}},\
  \bibinfo {pages} {889} (\bibinfo {year} {2017})}\BibitemShut {NoStop}%
\bibitem [{\citenamefont {Azzam}\ and\ \citenamefont
  {Kildishev}(2018{\natexlab{a}})}]{azzam2018sa}%
  \BibitemOpen
  \bibfield  {author} {\bibinfo {author} {\bibfnamefont {S.~I.}\ \bibnamefont
  {Azzam}}\ and\ \bibinfo {author} {\bibfnamefont {A.~V.}\ \bibnamefont
  {Kildishev}},\ }\href@noop {} {\bibfield  {journal} {\bibinfo  {journal}
  {Optical Materials Express}\ }\textbf {\bibinfo {volume} {8}},\ \bibinfo
  {pages} {3829} (\bibinfo {year} {2018}{\natexlab{a}})}\BibitemShut {NoStop}%
\bibitem [{\citenamefont {Azzam}\ and\ \citenamefont
  {Kildishev}(2018{\natexlab{b}})}]{azzam2018rsa}%
  \BibitemOpen
  \bibfield  {author} {\bibinfo {author} {\bibfnamefont {S.~I.}\ \bibnamefont
  {Azzam}}\ and\ \bibinfo {author} {\bibfnamefont {A.~V.}\ \bibnamefont
  {Kildishev}},\ }\href@noop {} {\bibfield  {journal} {\bibinfo  {journal}
  {Nanophotonics}\ }\textbf {\bibinfo {volume} {8}},\ \bibinfo {pages} {145}
  (\bibinfo {year} {2018}{\natexlab{b}})}\BibitemShut {NoStop}%
\bibitem [{Lum()}]{Lum}%
  \BibitemOpen
  \href@noop {} {}\bibinfo {note}
  {Lumericahttp://www.lumerical.com/tcad-products/fdtd/}\BibitemShut {NoStop}%
\bibitem [{\citenamefont {Yee}(1966)}]{yee1966fdtd}%
  \BibitemOpen
  \bibfield  {author} {\bibinfo {author} {\bibfnamefont {K.}~\bibnamefont
  {Yee}},\ }\href@noop {} {\bibfield  {journal} {\bibinfo  {journal} {IEEE
  Transactions on antennas and propagation}\ }\textbf {\bibinfo {volume}
  {14}},\ \bibinfo {pages} {302} (\bibinfo {year} {1966})}\BibitemShut
  {NoStop}%
\bibitem [{\citenamefont {Chang}\ and\ \citenamefont
  {Taflove}(2004)}]{taflove2004}%
  \BibitemOpen
  \bibfield  {author} {\bibinfo {author} {\bibfnamefont {S.}~\bibnamefont
  {Chang}}\ and\ \bibinfo {author} {\bibfnamefont {A.}~\bibnamefont
  {Taflove}},\ }\href@noop {} {\bibfield  {journal} {\bibinfo  {journal}
  {Optics Express}\ }\textbf {\bibinfo {volume} {12}},\ \bibinfo {pages} {3827}
  (\bibinfo {year} {2004})}\BibitemShut {NoStop}%
\end{thebibliography}%
\end{document}